\documentclass[a4paper,11pt]{article}

\usepackage{amsmath}
\usepackage{amsfonts}
\usepackage{amssymb}
\usepackage{amsthm}
\usepackage{caption}
\usepackage[english]{babel}
\usepackage{fontenc}
\usepackage{graphicx}
\usepackage{mathrsfs}
\usepackage{bbm}
\usepackage{array}
\usepackage{float}
\usepackage{enumerate}
\usepackage{rotating}
\usepackage{color}
\usepackage{hyperref}
\usepackage{fullpage}
\usepackage{centernot}
\usepackage{placeins}
\usepackage{natbib}

\usepackage{algorithm}
\usepackage{algorithmic}

\theoremstyle{plain}
\newtheorem{thm}{Theorem}

\theoremstyle{definition}

\theoremstyle{remark}

\newcommand{\ind}{\mathbbm{1}}
\newcommand{\pr}{\mathbb{P}}
\newcommand{\R}{\mathbb{R}}
\newcommand{\E}{\mathbb{E}}

\newcommand{\floor}[1]{\left\lfloor #1 \right\rfloor}

\newcommand{\tr}{\mathrm{tr}}
\newcommand{\sq}{\mathrm{sq}}

\DeclareMathOperator*{\argmin}{arg\,min}

\usepackage{authblk}

\title{High-dimensional regression with potential prior information on variable importance}

\author[1]{Benjamin G.\ Stokell\thanks{Supported by the Cantab Capital Institute for the Mathematics of Information.}}
\author[1]{Rajen D.\ Shah\thanks{Supported by EPSRC programme grant EP/N031938/1.}}
\affil[1]{University of Cambridge}

\begin{document}

\maketitle

\begin{abstract}
There are a variety of settings where vague prior information may be available on the importance of predictors in high-dimensional regression settings. Examples include ordering on the variables offered by their empirical variances (which is typically discarded through standardisation), the lag of predictors when fitting autoregressive models in time series settings, or the level of missingness of the variables. Whilst such  orderings may not match the true importance of variables, we argue that there is little to be lost, and potentially much to be gained, by using them. We propose a simple scheme involving fitting a sequence of models indicated by the ordering. We show that the computational cost for fitting all models when ridge regression is used is no more than for a single fit of ridge regression, and describe a strategy for Lasso regression that makes use of previous fits to greatly speed up fitting the entire sequence of models. We propose to select a final estimator by cross-validation and provide a general result on the quality of the best performing estimator on a test set selected from among a number $M$ of competing estimators in a high-dimensional linear regression setting. Our result requires no sparsity assumptions and shows that only a $\log M$ price is incurred compared to the unknown best estimator. We demonstrate the effectiveness of our approach when applied to missing or corrupted data, and time series settings. An R package is available on github.
\end{abstract}

\section{Introduction}
Regression with high-dimensional data is nowadays a routine task in large variety of application areas, ranging from genomic analysis and medicine to finance and industrial processes. Perhaps the most popular method is the Lasso \citep{tibshirani1996regression}, which given a response $Y \in \R^n$ and matrix of predictors $X \in \R^{n \times p}$ solves
\begin{equation} \label{eq:lasso}
\argmin_{b \in \R^p} \left\{ \frac{1}{2n} \|Y - X b\|_2^2 + \lambda\|b\|_1 \right\}.
\end{equation}
Whilst the one of the goals of the Lasso and the many related penalised regression procedures is to determine which variables are relevant, in many settings, some vague prior information on the relative importance of the variables may be available.

One instance of this concerns the scaling of variables so they have the same empirical variance before performing the Lasso optimisation \eqref{eq:lasso}. This practice is very common and is carried out by default in many software packages including the highly popular \texttt{glmnet} \citep{friedman2010regularization}. The rationale for this is to  ensure that all of the coefficients are treated in a balanced way; otherwise, coefficients corresponding to variables with large empirical variances will effectively experience very little shrinkage whereas those corresponding to variables with small variances will be penalised heavily. On the other hand, any information that may be encoded in the scale of the columns of $X$ is lost. For example, in a setting where one may expect measurement error to be distributed evenly over the variables, it is reasonable to suspect that variables with larger observed variance will be less corrupted by the error and hence contain more underlying signal. Perhaps as a result of this, it is also common to remove columns with the smallest variance as part of pre-processing, a step known in the machine learning community as applying a `low variance filter' (see for example \citet{filter2014seven,singh2017novel,abou2018comparative,langkun2020feature,saputra2018botnet,kalambe2020descriptor}).
This is however a somewhat crude way of using this potential information, and may eliminate important variables that happen to have a small variation, a risk which the practice of scaling variables aims to mitigate.


Another example of potential prior information being available concerns regression on lagged times series data when estimating autoregressive processes. It is natural to assumed that the importance of the historical data typically decreases with increasing lag, though seasonality considerations may also be incorporated into a final ordering among the predictors.
An a priori ordering or partial ordering on variables could arise for many other reasons, and in this paper we will focus on the more general question of how such information can be used to improve the final fit.

One approach to incorporating this information involves modifying the Lasso penalty term allowing individual tuning parameters $\lambda_j$ to be applied to each variable, $\sum_{j=1}^p \lambda_j | \beta_j|$, the idea being to place a smaller penalty on those coefficients believed to be more important.
Manual reweighting of the penalty terms is computationally very attractive since it is no more difficult to compute than the Lasso \eqref{eq:lasso}, for which there exist very fast and reliable algorithms that include this functionality \citep{friedman2010regularization}.
\citet{nardi2011autoregressive} study this approach in the autoregressive model setting.
Motivated primarily by the time-lagged regression example, \citet{tibshirani2016ordered} propose an alternative approach involving fitting the Lasso with a monotonicity constraint imposed on the coefficients with respect to the natural ordering.

Whilst these approaches can take great advantage of an informative ordering, one potential drawback  is that when the ordering is uninformative, their performance can suffer substantially. Indeed, the numerical results in \citet{tibshirani2016ordered} suggest that the practitioner pays a large price for supplying a randomised ordering. \citet{micchelli2010family} consider a general norm-based penalty framework that includes, as an example, a penalty applied to groups \citep{yuan2006model} which respect a specified ordering but are data-dependent. Such an approach, while less aggressive than a monotonicity constraint on the coefficients or a reweighting of the penalty, can still suffer if the ordering is unhelpful. In addition, both this and the approach of \citet{tibshirani2016ordered} are much more computationally involved than a regular Lasso regression, and less suited to the sort of large-scale settings we have in mind here.

In this work we propose to incorporate a potentially useful ordering over the variables in the following simple way, outlined in more detail in Section~\ref{sec:methodology}. We begin by fitting a model over the full set of variables and then proceed by fitting a sequence of nested submodels, each time removing a subset consisting of the least important variables according to our ordering. A final model is then selected by cross-validation. We show that when using ridge regression, by arranging computations appropriately, the cost for performing regressions on all $p$ nested subsets of variables given by the ordering is the same as that of a single fit on the full set of variables. We also describe a strategy for speeding up computations when the Lasso is used.

The statistical performance of our nested regression approach depends on cross-validation being able to select a good model from among the candidates given by the ordering, which in the case where the ordering is uninformative will most likely be close to the full model, but could be much smaller for an informative ordering. In Section~\ref{sec:generalresult} we give an oracle inequality bounding the performance of the coefficient estimate performing best on test data from among a sequence of estimates showing that the 
number of competing estimates only affects the performance logarithmically. 
In our context, these estimates will be derived from regressing on different subsets of variables, but the result may be of independent interest.

In Section~\ref{sec:experiments} we present the results of numerical experiments on simulated and real data that demonstrate the utility of our nested regression method. We study settings where order information may arise through knowledge of the degree of measurement error in the covariates, heterogeneous levels of missingness among the variables, and chronological ordering among variables derived from time series. We conclude with a discussion and proofs are deferred to the Appendix. An R package implementing our proposed methodology is available at \url{https://github.com/bgs25/OrderRegression}.

\section{Methodology} \label{sec:methodology}

In this section we present our nested regression framework for incorporating potential prior ordering information when fitting high-dimensional regression models. We first present our general approach before describing specific versions for Lasso and ridge regression in high-dimensional linear models.

\subsection{Using ordering information} \label{sec:general_method}
Consider a general regression problem, with response vector $Y \in \R^n$ and matrix of predictors $X \in \R^{n \times p}$, of the form
\begin{align}
\argmin_{b \in \R^p}\left\{ \ell(Y, Xb) + \mathcal{P}_\lambda(b)\right\}, \label{eq:regress_general}
\end{align}
where $\ell$ is some loss function and $\mathcal{P}_\lambda$ a penalty indexed by a regularisation parameter $\lambda$. Suppose that we have a reordering $\pi_1,\ldots,\pi_p$ of variables $1,\ldots,p$ with $\pi_j$ expected to be at least as important a predictor as $\pi_k$ when $k \geq j$. We then specify indices $p=j_1 > j_2 > \cdots > j_K=1$ from which we obtain a collection of nested subsets $S_1 \supset \cdots \supset S_K$ with $S_k = \{\pi_1,\ldots,\pi_{j_k}\}$. When $K \ll p$ we typically choose the $k_j$ such that sizes of the subsets decrease exponentially, so $|S_k|/|S_{k+1}| \approx |S_1|/|S_2|$ for $k = 1, \ldots, K - 1$, and $S_K = \{\pi_1\}$. This is partly motivated by the logarithmic cost associated with including extraneous variables in a Lasso regression. 

Let us introduce the notation $[j]=\{1,\ldots,j\}$ for $j \in \mathbb{N}$, and for any nonempty $S \subseteq [p]$, let $X_S \in \R^{n \times |S|}$ be the submatrix of $X$ consisting of the those columns indexed by $S$.

Given a grid of tuning parameters $\lambda_1 > \cdots > \lambda_L$, for each $k \in [K]$ and $l \in [L]$ we perform optimisation \eqref{eq:regress_general} with $\lambda = \lambda_l$ and regressing only on those variables indexed by $S_k$, i.e.\ with $X_{S_k}$ in place of $X$, to give a vector of coefficient estimates $\hat{\beta}^{k,l} \in \R^p$ with $\hat{\beta}^{k,l}_j = 0$ for all $j \notin S_k$. If our ordering were informative in that for large $k$, $S_k$ contained the set of important variables, regressions performed on $S_k$ would be unhampered by the potentially large number of unimportant variables in $[p] \setminus S_k$. For example, it may be the case that conditions on the design matrix such as the irrepresentable condition \citep{zhao2006model} or the compatibility condition \citep{van2009conditions} that guarantee the Lasso performs well are met by $X_{S_k}$ but not the full design matrix $X$.
Our new tuning parameter is the pair $(k,l)$, which we typically select by cross-validation. 

One potential issue with the approach outlined above is that particularly if $K$ is large, the computational burden of performing the $K \times L$ regressions may be large. We now explain how for both ridge regression and sparsity inducing penalties such as the Lasso, the computation can be organised so the cost is manageable even with large-scale data.

\subsection{Application to Lasso regression} \label{sec:lasso}
\begin{figure}
\centering
\begin{minipage}[b]{0.95\textwidth}
\includegraphics[width = \textwidth]{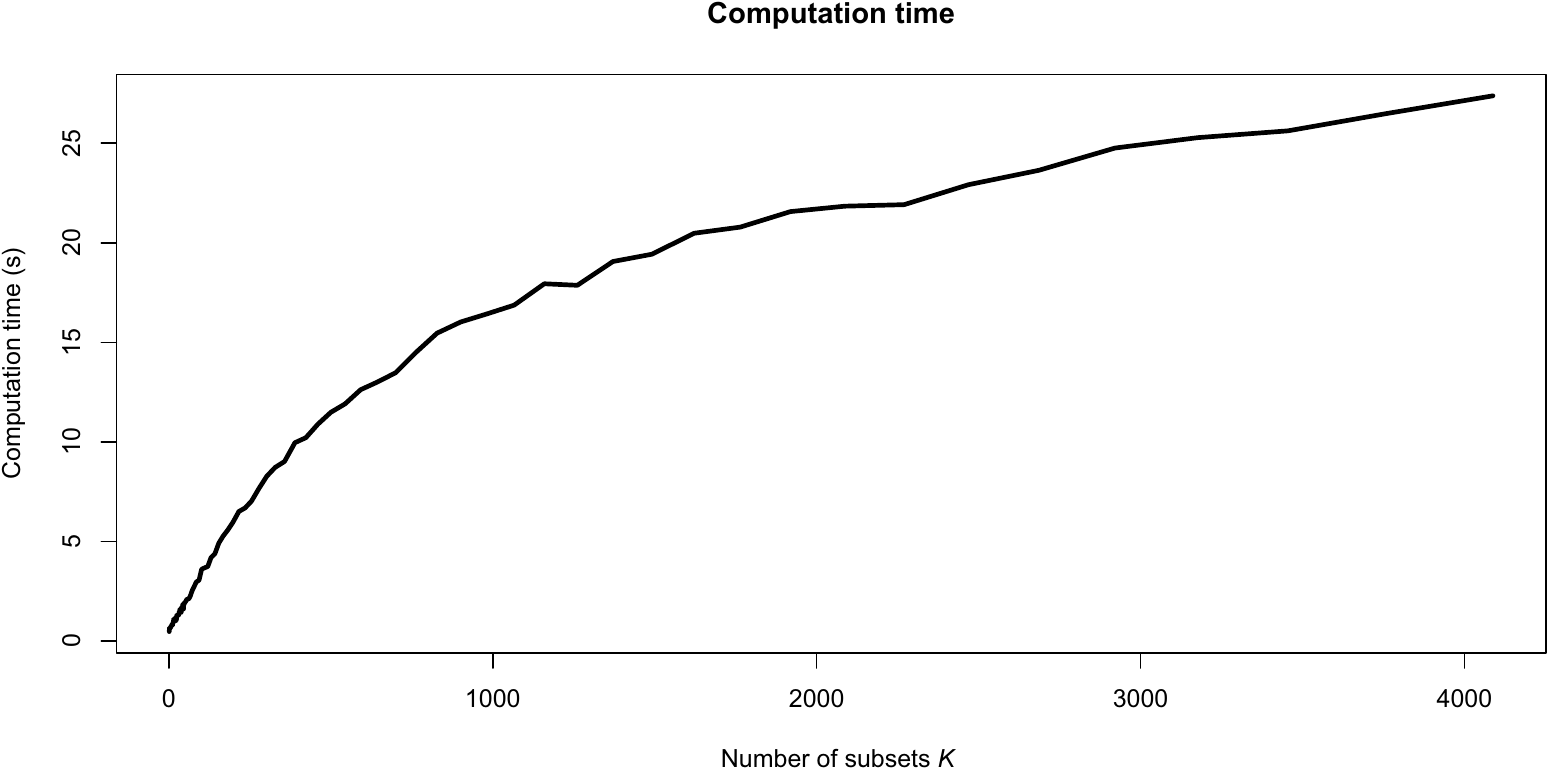}
\end{minipage}
\caption{Computation time of Algorithm~\ref{alg:lasso} as a function of the number of subsets $K$ when applied to the \emph{Riboflavin} dataset \citep{dezeure2015high}, where $n = 71$ and $p = 4088$.
\label{fig:comptime}}
\end{figure}

We now consider nested regression using the Lasso \eqref{eq:lasso} as our base regression procedure.
Similarly to how warm starts greatly speed up the computation of a Lasso solution path compared to separately computing coefficient estimates at each $\lambda_l$ \citep{friedman2010regularization}, we can utilise $\{\hat{\beta}^{k,l}\}_{l=1}^L$ to speed up computation of $\{\hat{\beta}^{k+1,l}\}_{l=1}^L$. The key observation is that if for a given $(k,l)$, $\hat{\beta}^{k,l}_j = 0$ for all $j \notin S_{k+1}$, then $\hat{\beta}^{k+1,l} = \hat{\beta}^{k,l}$, so no further computation is required. The finer the grid (i.e.\ the larger $K$ is), the more computation can be skipped and thus the greater the relative gains of using this are (as demonstrated in Figure~\ref{fig:comptime}). This strategy can also be used for other penalised regression approaches that yield sparse solutions, such as the relaxed Lasso \citep{meinshausen2007relaxed}, the adaptive Lasso \citep{zou2006adaptive} and MCP \citep{zhang2010nearly}, or regression procedures using sparsity inducing penalty functions with losses other than squared error loss. 

In the case of the Lasso, a further speedup may be realised by recognising that the bulk of the computation in the solution path occurs when computing the tail of path, i.e., those solutions with small values of the tuning parameter. In many settings, such solutions will not be needed as they will not be selected by cross-validation. The square-root Lasso offers an approach to eliminate these solutions without needing to compute them. The square-root Lasso \citep{Belloni2011, sun2012scaled} is given by
\begin{equation} \label{eq:sqrt_lasso}
 \hat{\beta}_{\sq}(\lambda_{\sq}) \in \argmin_{b \in \R^p} \left\{\|Y - X b\|_2/\sqrt{n} + \lambda_{\sq} \|b\|_1\right\}.
\end{equation}
By comparing the KKT conditions of the optimisation above and those of the Lasso, it may be shown that the solution paths of the Lasso \eqref{eq:lasso} and the square-root Lasso are identical, but are parametrised differently. Specifically, writing $\hat{\beta}(\lambda)$ for a Lasso solution \eqref{eq:lasso} with tuning parameter $\lambda$ and $\hat{\sigma}(b) := \|Y - X b\|_2 / \sqrt{n}$ for $b \in \R^p$, we have that $\hat{\beta}(\lambda)$ is in fact a square-root Lasso solution with tuning parameter $\lambda_{\sq} = \lambda / \hat{\sigma}(\hat{\beta}(\lambda))$, provided $\hat{\sigma}(\hat{\beta}(\lambda)) > 0$.
Conversely, any square-root Lasso solution $\hat{\beta}_{\sq}(\lambda_{\sq})$ is also a Lasso solution with $\hat{\beta}(\lambda)$ with $\lambda = \lambda_{\sq} \hat{\sigma}(\hat{\beta}_{\sq}(\lambda_{\sq}))$, provided $\hat{\sigma}(\hat{\beta}_{\sq}(\lambda_{\sq}))>0$.
Furthermore, $\lambda \mapsto \lambda / \hat{\sigma}(\hat{\beta}(\lambda))$ is a well-defined non-decreasing function; 
%
%
%
see \citet[Sec.~B]{shah2019double} for a derivation.

One key advantage of the square-root Lasso is that there exist theoretically rate-optimal choices of the tuning parameter, such as $1.1\sqrt{2\log (p)/n}$, that do not depend on the unknown variance of the noise. In practice however, such theoretically motivated choices tend to be conservative in that they are too large, and are typically out-performed by cross-validation at least from the perspective of prediction error. However 
they
can nevertheless be helpful as a criterion for terminating the computation of the Lasso solution path at a $\lambda_l > \lambda_L$.
Writing $\lambda(\lambda_{\sq}) = \lambda_{\sq} \hat{\sigma}(\hat{\beta}_{\sq}(\lambda_{\sq}))$ for a maximal $\hat{\sigma}(\hat{\beta}_{\sq}(\lambda_{\sq}))$ for instance\footnote{The quantity $\hat{\sigma}(\hat{\beta}_{\sq}(\lambda_{\sq}))$ defined by \eqref{eq:sqrt_lasso} may not be unique as $\hat{\beta}_{\sq}(\lambda_{\sq})$ may not be unique.},
if we take $\lambda_{\sq}$ to be say half of a theoretically optimal choice of tuning parameter (in our experiments we used $0.5$ times the choice suggested in \citet{sun2013sparse}), cross-validation is unlikely to select any $\lambda_l < \lambda(\lambda_{\sq})$, and so such solutions need not be computed.

Applying the two measures outlined above yields substantial reduction in the effort required to compute a full set of solution paths across all nested subsets, as shown in Figure~\ref{fig:comptime}.
Algorithm~\ref{alg:lasso} below describes how our nested regression approach is implemented. The sets $A(k, l)$ record which variables have been included in the solution path for variable set $S_k$ at $\lambda = \lambda_l$.

\begin{algorithm}
\caption{Algorithm for nested Lasso regression} \label{alg:lasso}
	\begin{algorithmic}[1]
		\REQUIRE{Matrix of predictors $X \in \R^{n \times p}$, response $Y \in \R^n$, nested sets of variables $S_1 \supset \cdots \supset S_K$, tuning parameter grid $\lambda_1 > \cdots > \lambda_L$, square-root Lasso parameter for early stopping $\lambda_{\sq} > 0$ (we use $0.5$ times the choice of  \citet{sun2013sparse})}
		\STATE 
		\STATE Set $A(0, l) = \{ 1, \ldots, p + 1 \}$ for $l = 1, \ldots, L $ and $\hat{\beta}^{k, 0} = 0$ for $k = 1, \ldots, K$ 
		\FOR{$k = 1, \ldots, K$}
		\FOR{$l = 1, \ldots, L$}
		\IF{$A(k - 1,l) \subseteq S_k$}
		\STATE Set $\hat{\beta}^{k, l} = \hat{\beta}^{k-1,l}$ and $A(k, l) = A(k-1, l)$
		\ELSIF{$\| Y - X \hat{\beta}^{k-1,l} \|_2 / \lambda_l \leq \sqrt{n} / \lambda_{\sq}$}
		\STATE Compute $\hat{\beta}^{k, l} = \argmin_{\beta : \beta_{S_k^c} = 0} \{\frac{1}{2n} \| Y - X \beta \|_2^2 + \lambda_l \| \beta \|_1\}$ using $\hat{\beta}^{k, l-1}$ as an initial estimate
		\STATE Set $A(k, l) = \{j : \hat{\beta}^{k,l} \neq 0\}$
		\ELSE
		\STATE Set $\hat{\beta}^{k,l} = \hat{\beta}^{k,l-1}$ and $A(k,l) = A(k,l-1)$
		\ENDIF
		\ENDFOR
		\ENDFOR
		\ENSURE{$\{ \hat{\beta}^{k,l} \colon k \in \{1, \ldots, K \}, l \in \{ 1, \ldots, L\} \}$}
	\end{algorithmic}
\end{algorithm}

\subsection{Application to ridge regression} \label{sec:ridge}
We now turn to the case where ridge regression \citep{hoerl1970ridge} is the base procedure in our nested regression approach. Given a tuning parameter $\lambda > 0$, the solution on the full set of variables is given by
\begin{equation} \label{eq:ridge}
(X^TX + \lambda I )^{-1} X^T Y = X^T (XX^T + \lambda I)^{-1} Y,
\end{equation}
with the right-hand side computable in $O(n^2p)$ operations in the case $n \gg p$.
We will consider the specific case where $K=p$, and so without loss of generality, we may assume $S_k = \{k,\ldots,p\}$.
We will show that for a fixed $\lambda$ and test set $Z \in \R^{n' \times p}$, computation of the full collection of test set predictions for ridge regression solutions across all $S_1,\ldots,S_p$ also requires only $O(n^2p)$ operations, provided $n' \leq n$ (which we will from now assume). Applying this across a grid of $L$ tuning parameter values then gives a total cost of $O(Ln^2p)$ for determining the $(k,l)$ pair by cross-validation, when the number of folds is constant.

To achieve this, it will be convenient to compute predictions corresponding to $S_p$ first, and then use this to obtain the predictions for $S_{p-1}$, and hence $S_{p-2}$ and so on, as we now explain. Let us fix $k \leq p$.
The out-of-sample predictions corresponding to \eqref{eq:ridge} applied to matrix of predictors $X_{S_{k+1}} := X^{(1)}$ and test data $Z_{S_{k+1}} := Z^{(1)}$ are given by
\begin{equation} \label{eq:ridge_pred}
 Z^{(1)}(X^{(1)})^T \{X^{(1)}(X^{(1)})^T + \lambda I\}^{-1} Y.
\end{equation}
Let us suppose that $Z^{(1)}\{X^{(1)}\}^T \in \R^{n' \times n}$ and $A:=(X^{(1)}(X^{(1)})^T + \lambda I)^{-1} \in \R^{n\times n}$ have been computed; note these are required in calculating \eqref{eq:ridge_pred} which then needs only an additional $O(n^2)$ operations to compute. 
We now explain how 
the corresponding quantities for $X^{(2)} := (X_k,X^{(1)})=X_{S_k}$ and $Z^{(2)} := (Z_k, Z^{(1)})=Z_{S_k}$ may then by retrieved in $O(n^2)$ steps, where $X_k \in \R^n$ and $Z_k \in \R^{n'}$ are the $k$th columns of $X$ and $Z$ respectively. First observe that
\[
Z^{(2)}(X^{(2)})^T = Z^{(1)}(X^{(1)})^T + Z_kX_k^T,
\]
so $Z^{(2)}(X^{(2)})^T$ may be formed using $O(nn')$ operations. Next applying the Sherman--Morrison--Woodbury rank one update formula, we see that
\begin{align*}
	\{X^{(2)}(X^{(2)})^T + \lambda I\}^{-1} &=  \{X^{(1)}(X^{(1)})^T + X_kX_k^T + \lambda I\}^{-1} \\
	&= \{X^{(1)}(X^{(1)})^T + \lambda I\}^{-1} - \frac{AX_k X_k^T A}{1 + X_k^T A X_k}, 
\end{align*}
with right-hand side easily obtained from $A$ in $O(n^2)$ computations. We thus see that the entire set of solutions may be computed using $O(pn^2)$ operations.

In practice we compute the full model first using the singular value decomposition of $X$ to obtain the full solution path for this model, with an analogous rank-one-update to that above applied to progress to smaller models.

To illustrate the speed with which this enables the models to be computed, we again used the \emph{Riboflavin} dataset (see Section~\ref{sec:riboprost}). We trained the models on $50$ observations and computed predictions for the remaining $21$ across a sequence of $100$ tuning parameter values and $K=p$ variable subsets.
Thus, the total number tuning parameter pairs considered was $408\,800$; the full computation for all of these on a standard laptop took under 100 seconds. 
We obtained an estimated mean squared prediction error of $0.318$ for the model fitted only on the top $169$ nodes, compared to $0.399$ for the full model.

\section{Theory} \label{sec:generalresult}
Our proposed nested regression scheme for incorporating potential prior information in high-dimensional regression involves consideration of a number of different estimators corresponding to different tuning parameter values and subsets of variables. The statistical performance relies on cross-validation being able to select a good estimator from among these. In this section we study this general problem in the context of the high-dimensional linear model,
\begin{equation} \label{eq:lin_mod}
	Y = X \beta + \varepsilon,
\end{equation}
for a simplified form of cross-validation. Specifically, we suppose that we have available candidate estimators $\hat{\beta}^{(1)}, \ldots, \hat{\beta}^{(M)} \in \R^p$ trained on data $(Y^{\tr}, X^{\tr})$, independent of test data $(Y, X) \in \R^n \times \R^{n \times p}$, which we will treat as fixed. In the context of the previous section, $M$ would be $L K$. With these we compute
\begin{align}
	\hat{\beta} \in \argmin_{b\in \{ \hat{\beta}^{(1)}, \ldots, \hat{\beta}^{(M)} \}} \|Y - X b \|_2^2. \label{eq:betahat_def}
\end{align}
The candidate estimators could, for example, have been obtained via Lasso or ridge regression; however we consider a more general setting that allows for arbitrary candidate estimators.

Suppose that the rows of $X$ are i.i.d.\ with mean zero and covariance $\Sigma \in \R^{p \times p}$. We compare the performance of $\hat{\beta}$ to that of the unknown best estimator among $ \hat{\beta}^{(1)}, \ldots, \hat{\beta}^{(M)}$ defined by
\begin{align}
	\hat{\beta}^* \in \argmin_{b \in \{\hat{\beta}^{(1)},\ldots,\hat{\beta}^{(M)}\}} (\beta - b)^T\Sigma(\beta-b). \label{eq:oracle_def}
\end{align}
While a large $M$ can be expected to result in $\hat{\beta}^*$ and $\beta$ being closer, naturally it will result in a larger discrepancy in the performance of $\hat{\beta}^*$ and $\hat{\beta}$. 
Theorem~\ref{thm:mainresult} however indicates that the price to pay is only logarithmic in $M$. The implication for our nested regression scheme is that $K$ and $L$ can be chosen to be relatively large subject to computational constraints; while they may not be optimal statistically, they will not be too far off.
\begin{thm} \label{thm:mainresult}
Suppose that $X = W \Sigma^{1/2}$ where $W \in \R^{n \times p}$ has independent mean-zero sub-Gaussian entries with variance proxy $\nu^2$. Suppose linear model \eqref{eq:lin_mod} holds with the components of $\varepsilon$ mean-zero, independent and sub-Gaussian with variance proxy $\sigma^2$. Then for any $c_1, c_2 >0$ with $c_1<n / \log M - 1$ , we have that with probability at least $1 - 2M^{-c_1} - 2M^{-c_2}$,
\begin{align}
		\| \Sigma^{1/2} ( \hat{\beta} - \beta ) \|_2 &\leq \frac{1 + \Psi}{1 - \Psi}  \| \Sigma^{1/2}(\hat{\beta}^* - \beta ) \|_2  + \frac{1}{1 - \Psi}2\sqrt{2}\sigma \sqrt{1 + c_2} \sqrt{\frac{\log M}{n}}, \label{eq:generalresult} 
		\end{align}
		where $\Psi = 2\sqrt{2}\nu(1 + c_1)^{1/4} \{(\log M) / n \}^{1/4}$.
\end{thm}
Note that no sparsity assumptions on $\beta$ are required, and $p$ can be arbitrarily large compared to $n$. In addition, no requirements are placed on the quality of the candidate estimators which can be arbitrarily good or bad, though this will be reflected in $\hat{\beta}^*$ of course.

In practice we would use cross-validation rather than a single training and test split of the original data as considered in our setup here. However existing results on cross-validation are not available in the sort of generality we consider here; in particular they do require sparsity assumptions and are tied to particular estimators such as the Lasso  \citep{feng2019restricted,chetverikov2021cross}.

We remark that another option for comparing the different models output could be AIC or BIC, which have the advantage of not requiring test data. However, such information criteria would not have a way of discriminating between the complexity a model fitted on $r$ variables selected out of set $S_k \subset [p]$, or one selected out of the full set of $p$ variables $S_K$. On the other hand, cross-validation implicitly provides a way of doing this: the model selected out of the full set of variables may have a greater degree of over-fitting which would be exposed by performance on the test set.

\section{Numerical experiments} \label{sec:experiments}
In this section we explore the properties of our nested regression approach in a range of scenarios, using both simulated and real data. In Section~\ref{sec:ordersims} we consider different levels of `informativeness' in the orderings, and the effect that this has on the prediction error of the final model. 
Section~\ref{sec:riboprost} explores the effect of varying the number of subsets $K$ using two real datasets. 
Sections~\ref{sec:corrupted}~and~\ref{sec:missing} again use real data, this time exploring how our approach can be used with missing or corrupted data. 
Lastly, in Section~\ref{sec:realdata} we run our method on a dataset to predict avocado prices, illustrating the use of our approach in a time series context. For our nested regression approach, we use the Lasso with $5$-fold cross-validation and the default grid of 100 tuning parameters as chosen by \texttt{glmnet} on the full model $S_1$.

\subsection{Quality of ordering} \label{sec:ordersims}
\begin{figure}[h!]
\centering
\includegraphics[width=\textwidth]{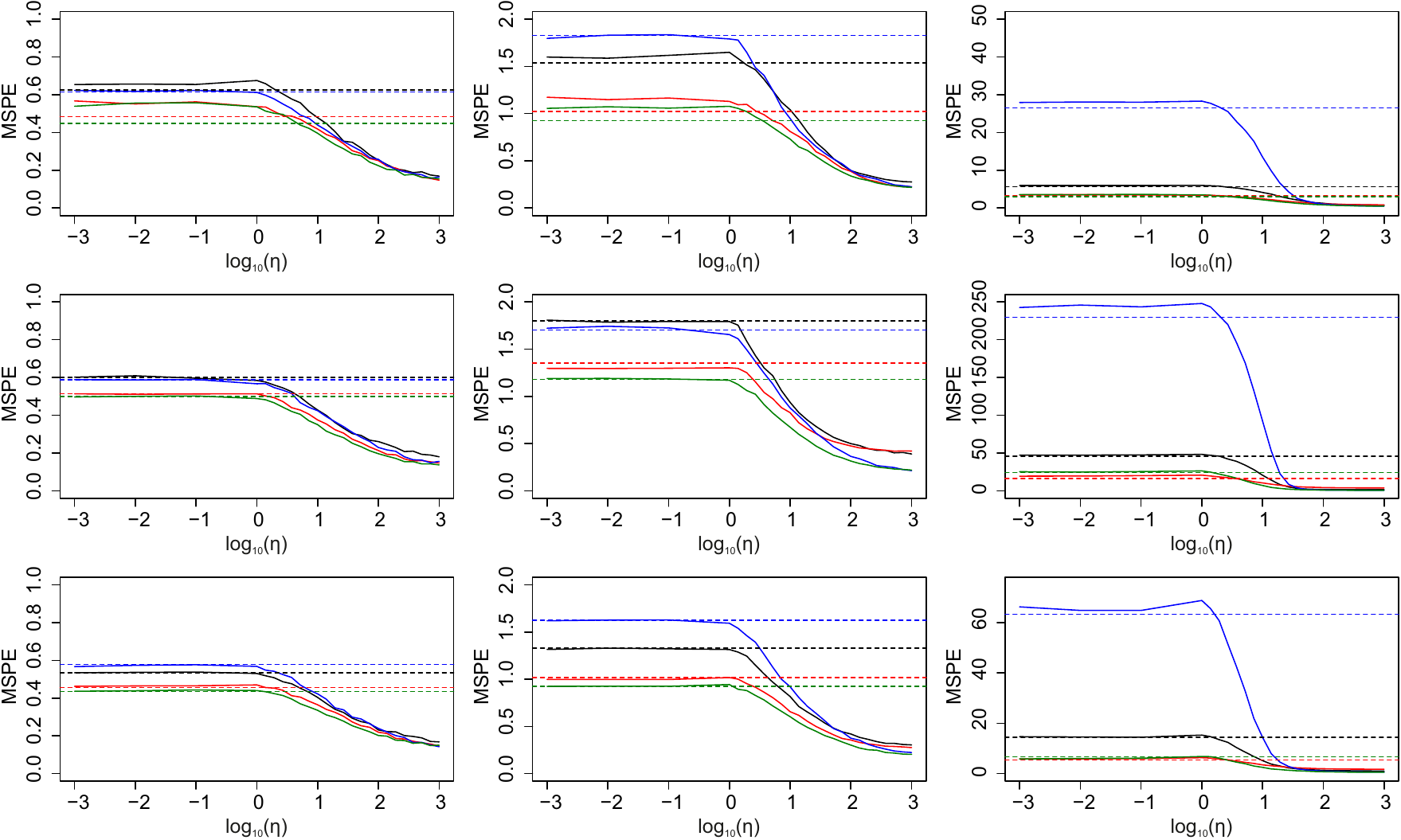}

\caption{Mean squared prediction errors (MSPE) of our nested regression method as the logarithm of the probability ratio $\eta$ varies: 0 relates to a neutral choice of ordering, larger and negative means a more adversarially bad choice, larger and positive means a  more informative choice. Left--Right: 5, 10, 25 signal variables; Top--Bottom: 0.5, 1.5, $U[0,2]$ signal coefficients. Colours for settings are 1.\ black, 2.\ red, 3.\ blue, 4.\ green. The dotted lines correspond to the errors achieved by the standard approach on the same data.
\label{fig:ordererrors}}	
\end{figure}
In order to see the effect of different variable orderings on the performance of the model, we sample orderings weighted by a vector $\rho \in [0,1]^p$ of probabilities using the \texttt{sample} function in \texttt{R} \citep{R} supplied with the $\rho$ as the \texttt{prob} argument. This uses so-called size-biased sampling as described in \citet{pitman2015size}.
For example, a neutral (or uninformative) ordering would have this vector as $( 1/p, \ldots, 1/p)$ (where $p$ is the number of variables), as all permutations are equally likely.

After a model has been constructed, with a true support $S$, we specify the $j$th entry of $\rho$ for $j = 1, \ldots, p$ by
\begin{align*}
	\rho_j =  \begin{cases}
		\eta / (p + (\eta  -1)|S|) &\text{ if } j \in S \\
		1 / (p + (\eta-1)|S|) &\text{ otherwise, }
	\end{cases}
\end{align*}
where $\eta > 0$ is the `probability ratio'.
A choice of $\eta > 1$ means that the ordering is more likely to favour signal variables (meaning the ordering is likely to be useful), whereas $\eta <1$ means the ordering will prefer non-signal variables (meaning that the ordering will be actively unhelpful). The vector $\rho$ is used as the weight vector for sampling a permutation which was then used as the ordering.

We consider a linear model \eqref{eq:lin_mod} with design matrices sampled with $n = 100$ and $p = 1000$, with i.i.d. mean-zero Gaussian rows with covariance matrix $\Sigma$. Tests were run with four different choices of $\Sigma$:
\begin{enumerate}
  \item  $\Sigma_{jk} = \ind_{\{ j = k \} }$
  \item	$\Sigma_{jk} = 0.9^{|j - k|}$
  \item	$(\Sigma^{-1})_{jk} = 0.4^{|j - k| / 5}$ ($\approx 0.833^{|j - k|}$)
  \item	$\Sigma_{jk} = 0.5 + 0.5\ind_{\{ j = k \}}$.
\end{enumerate}
Three different sparsities for $\beta$ were used: 5, 10, and 25 variables, with the variables with non-zero coefficients selected uniformly at random. Three regimes for populating the non-zero entries in $\beta$ were used: two constant (0.5 and 1.5), and one random, where coefficients are drawn as independent $U(0,2)$ random variables. We fixed the number of subsets $K$ to be $100$; in these examples, the results were relatively insensitive to the choice of $K$ provided it was sufficiently large, a phenomenon expected to hold more generally given the result of Theorem~\ref{thm:mainresult}.
The tests were repeated 500 times with \texttt{glmnet} used to perform the regressions in the case with $K=1$.

It is interesting to note in Figure~\ref{fig:ordererrors} that an adversarially bad ordering does not give rise to any worse performance than a neutral one. 
There is reason to believe that this should be preferable: in a setting where an ordering is not actively helpful (i.e.\ it is either neutral or adversarially poor) then we wish for our procedure to select the full model, $S_1$. If the ordering is sufficiently bad that the increase in loss for the submodels is larger than the variance of the test error, there is a greater chance that $S_1$ will be selected.

\subsection{Riboflavin and Prostate data} \label{sec:riboprost}
\begin{figure}[h!]
\centering
\begin{minipage}[b]{0.49\textwidth}
\includegraphics[width=\textwidth]{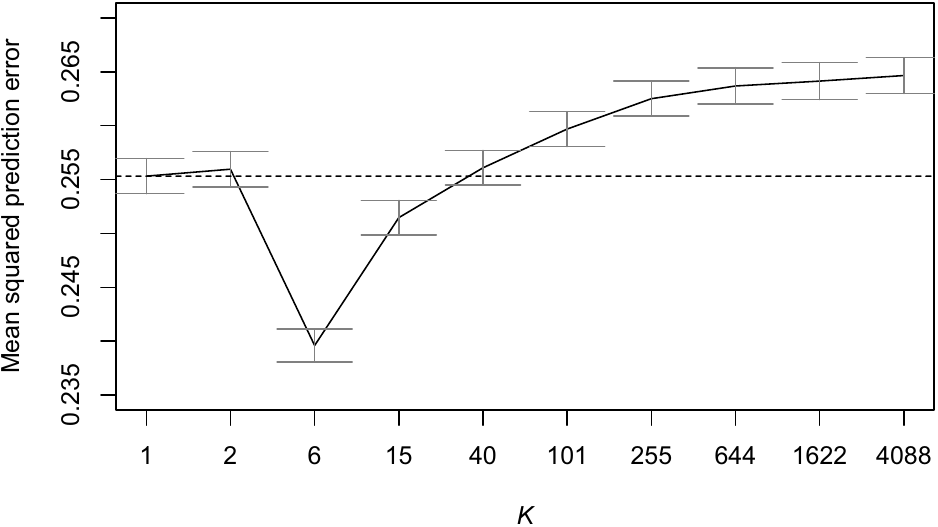}
\end{minipage}
\begin{minipage}[b]{0.49\textwidth}
\includegraphics[width=\textwidth]{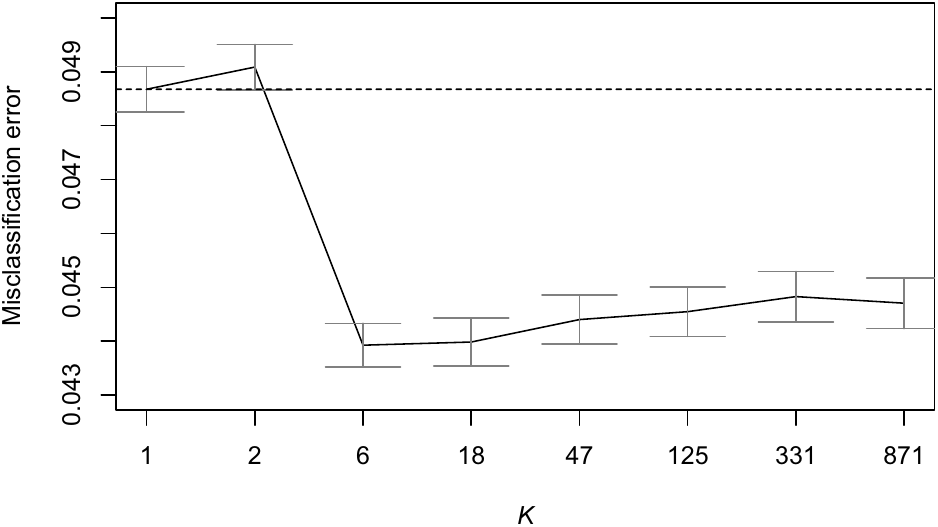}
\end{minipage}
\caption{Average prediction errors for order regressions on riboflavin (left) and prostate (right) datasets for different $K$, with squared error and misclassification losses respectively. The error bar is $\pm2$ standard deviations and the dotted line depicts the prediction error of a standard $\ell_1$-penalised regression.
\label{fig:riboflavinprostate}}	
\end{figure}
Tests were performed on the \emph{riboflavin} dataset (available in R package \texttt{hdi} \citep{dezeure2015high}; $n=71$, $p=4088$) and the \emph{prostate} dataset (available in R package \texttt{spls} \citep{chun2010sparse}; $n=102$, $p=6033$) to explore the performance improvements attained by using our nested regession approach.
With the former dataset, which has a continuous response, we used a Lasso regression with square-error loss, and with the latter, which has a binary response, we used an $\ell_1$-penalised logistic regression.
Prediction error was estimated by cross-validation error with 5-folds; within each of the folds cross-validation was also used to select the model. A range of values of $K$ were used for each dataset shown by the ticks on the plots in Figure~\ref{fig:riboflavinprostate}. 
The ordering used for both of these was the one induced by the scales of the columns in the matrices of predictors. As mentioned in the introduction, this order is often used more directly in a low variance filtering step, particularly with gene expression data \citep{singh2017novel}.

We see in both of these examples that the prediction error appears to improve after using a nested regression approach instead of ordinary Lasso models (which is equivalent to a $K= 1$).
With the both the riboflavin data and the prostate data, the error then increases slowly as $K$ increases. This behaviour is to be expected from \eqref{eq:generalresult} where the the bound on the right-hand-side is typically decreasing in $M=LK$, whereas the final term is increasing in $M$.


Only 8 values of $K$ were used for the prostate data due to computational constraints, as for logistic regression models we used a simple loop over $\ell_1$-penalised logistic regressions rather than the approach described in Section~\ref{sec:lasso}.
Each test was repeated 2000 times which different splits used in the cross-validation schemes in each run.

\subsection{Corrupted data} \label{sec:corrupted}
Here we consider a setting where an ordering among the variables arises through corruption of the entries of the design matrix.
We used the `muscle-skeletal' dataset from the GTEx project\footnote{\texttt{https://gtexportal.org}}, which has 491 rows and $14\,713$ columns, as our uncorrupted dataset. We preprocessed the data as in \citet{shah2020right} by removing the effect of measured and estimated confounders. 
We took as a response variable a column randomly selected (anew in each run) from the matrix, meaning that for our experiment $n = 491$ and $p = 14\, 712$.

We then artificially corrupted the data by independently, for each $j$ replacing with probability $\rho_j$ each entry in $j$th column of the matrix of predictors by independent standard Gaussian random variables.
In order to obtain an ordering we assume knowledge of the ranking of the $\rho_j$. In practice, if this is not known then an estimate can still be useful, as even if a practitioner has only a very vague notion of which are more likely to be corrupted, we see in Section~\ref{sec:ordersims} that it can still be beneficial to use such an ordering.

 \begin{figure}
\centering
\includegraphics[width=\textwidth]{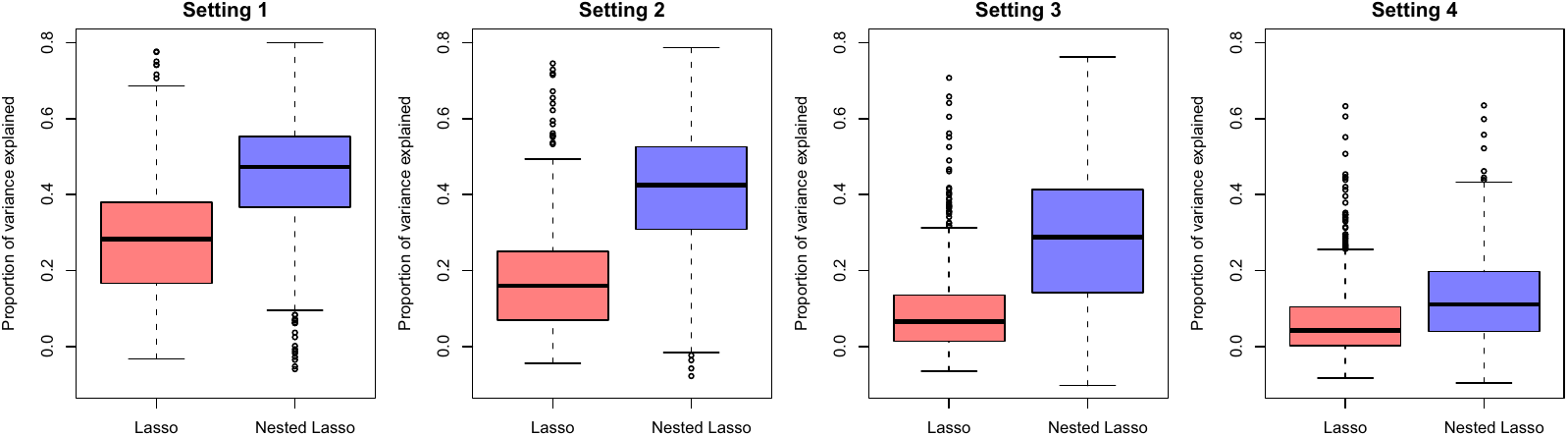}
\caption{Proportion of variance explained (larger is better)
	by the Lasso and our nested regression scheme incorporating ordering information for each of the corruption regimes.
\label{fig:corrupteddatabox}}	
\end{figure}

We tested performance in four settings, each with a different vector $\rho$ controlling the corruption probabilities of the variables. We constructed $\rho_1, \ldots, \rho_p$ for each of the four settings as follows (before randomising the order of $\rho_1,\ldots, \rho_p$):
\begin{enumerate}
  \item $\rho_j = 0.5$ for $j = 1, \ldots, \floor{ 0.2 p }$ and 0 otherwise
    \item $\rho_j = 0.5$ for $j = 1, \ldots, \floor{ 0.5 p }$ and 0 otherwise
  \item $\rho_j = 0.5$ for $j = 1, \ldots, \floor{ 0.8 p }$ and 0 otherwise
  \item $\rho_j = \min \{ 0.95,(j-1)/ p \}$
\end{enumerate}

The data were split into five folds; for each of these a model was fitted on the complementary four with entries corrupted according the settings above. 
These models were themselves tuned using five-fold cross-validation.
We set $K=25$ for all experiments, which were repeated 2000 times.

\subsection{Heterogeneous missing data} \label{sec:missing}
Here we consider the missing data setting, where in a given design matrix $X$, each entry $X_{ij}$ is missing independently with probability $\rho_j$. In contrast to the previous section, here it is observed exactly which entries are missing. This means that the ranking of the variables by their overall missingness is known, with no additional knowledge assumed.

Data are missing homogeneously in the case where $\rho_j \equiv \rho$, i.e.\ the entries in $X$ are all missing with equal probability.
In this case, the probability $\rho$ can be estimated and there are well-studied methods for computing Lasso solutions, such as discussed in \citet{loh2012high}.
However, the setting we consider here includes heterogeneous missing data.
Within high-dimensional statistics there are methods that accommodate heterogeneous missingness
in principal components analysis \citep{zhu2019high} and 
in regression problems \citep{rosenbaum2013improved, datta2017cocolasso}.

In order to implement our approach in this setting, we first observe that the Lasso objective can be written as
\begin{align}
\frac{1}{2n} Y^T Y - \frac{1}{n} b^T X^T Y + \frac{1}{2n}b^T X^T X b + \lambda \|b \|_1 ,\label{eq:cov_objective}	
\end{align}
which depends on $X$ only through the vector $X^T Y /n$ and matrix $X^T X /n$. Indeed, this is the starting point of existing approaches for performing regression on high-dimensional data with missing entries \citep{loh2012high, rosenbaum2013improved, datta2017cocolasso}.
In the case where $X$ has some missing entries, the above quantities can be estimated by:
\begin{align}
\hat{\Gamma}_{jk} &= \tilde{X}_j^T \tilde{X}_k / |\{ i : X_{ij} \text{ and } X_{ik} \text{ not missing}\}| \label{eq:missingGam}\\
\hat{\gamma}_{j} &= \tilde{X}_j^TY / |\{i : X_{ij}\text{ not missing }\}|, \label{eq:missinggam}
\end{align}
where $\tilde{X}_{ij} = X_{ij}$ if $X_{ij}$ not missing, and 0 otherwise. 
These quantities can be substituted into the update steps in computing solutions to the following surrogate  objective function
\begin{align}
\hat{\beta} \in \argmin_{b \in \R^p} \left\{- b^T\hat{\gamma}_j + \frac{1}{2} b^T \hat{\Gamma}b + \lambda \| b \|_1\right\}. \label{eq:cov_objective2}	
\end{align}
Note that $\hat{\Gamma}$ is not in general positive semidefinite, which would be needed in order for \eqref{eq:cov_objective2} to have a finite minimum. 
To overcome this difficulty, \citet{datta2017cocolasso} suggest projecting $\hat{\Gamma}$  onto the cone of positive semi-definite matrices via $\argmin_{\Gamma \in \mathbf{S}^p_{+}}\|\Gamma - \hat{\Gamma}\|_{\infty}$, which restores the convexity of the problem.
Here we will use a simpler approach replacing
 $\hat{\Gamma}$ with $\hat{\Gamma}_{\text{psd}} := \hat{\Gamma} + \Lambda_{\text{min}}(\hat{\Gamma})I_p$, where $\Lambda_{\text{min}}(\hat{\Gamma})$ is the smallest eigenvalue of $\hat{\Gamma}$.
This is equivalent to the addition of a fixed ridge penalty term in the objective function
\begin{align}
\hat{\beta} &\in \argmin_{\beta \in \R^p} \left\{- \beta^T\hat{\gamma}_j + \frac{1}{2} \beta^T \hat{\Gamma}_{\text{psd}} \beta + \lambda \| \beta \|_1 \right\} \notag \\
&= \argmin_{\beta \in \R^p} \left\{- \beta^T\hat{\gamma}_j + \frac{1}{2} \beta^T \hat{\Gamma} \beta + \lambda \| \beta \|_1 + \frac{1}{2}\Lambda_{\text{min}}(\hat{\Gamma})\|\beta\|_2^2 \right\}. \label{eq:cov_objective3}	
\end{align}
Tests were performed on the same `muscle-skeletal' dataset as in Section~\ref{sec:corrupted}, again randomly selecting a variable in each replicate to use as the response. 
Three missing data regimes were used, each determined by a vector $\rho \in [0,1]^{p}$ specifying the independent missingness probability of each variable:
\begin{enumerate}
  \item $\rho_j = 0.25$ for all $j$,
  \item  $\rho_j = (j-1)/3p$, then randomising the order of $\rho$,
  \item $\rho_j = 0.3$ for $j = 1, \ldots, \floor{ 0.5p }$ and 0 otherwise, then randomising the order of $\rho$.
  \end{enumerate}

 \begin{figure}
\centering
\includegraphics[width=\textwidth]{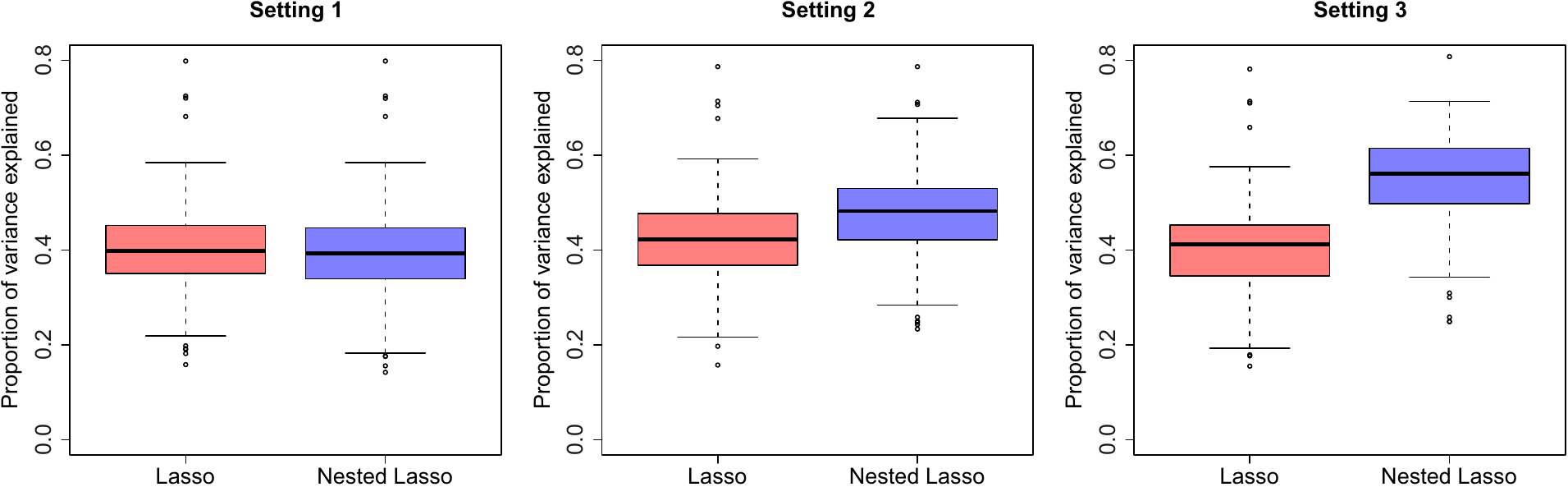}
\caption{Proportion of variance explained (larger is better)
	by the Lasso and our nested regression scheme incorporating ordering information for each of the missing data regimes. 
\label{fig:missingdatabox}}	
\end{figure}

 The data were split into five folds; for each of these a model was fitted on the complementary four with missing entries according to the settings described. 
 These models were themselves tuned using five-fold cross-validation, with scores computed using the estimates $\hat{\Gamma}$ and $\hat{\gamma}$. 
 Experiments were repeated 250 times.

Figure~\ref{fig:missingdatabox} displays the proportion of variance explained by the both the vanilla Lasso and by our approach with $K=25$ in each of the above settings.
 As expected, there is no improvement in Setting 1 from using a nested regression approach, as the missingness is homogeneous so the ordering will be uniformly random.
 In the other two settings, using our approach with the knowledge of which variables are missing more frequently allows us to fit models that provide better predictions.

\subsection{Avocado data} \label{sec:realdata}
Here we study historical data on avocado prices and sales volume in multiple US markets available on \emph{Kaggle} \citep{avocadokaggle}.  
For this experiment we consider predicting the price using an autoregressive model, and use the 53 markets for which full weekly price data is available for both `conventional' and `organic' varieties from the beginning of January 2015 to the end of March 2018.

A design matrix was compiled using all 106 time series, using the previous 52 values for each of one, thus resulting in a $5512$ ($ = 52 \times 106$)-dimensional model, with 117 observations. 
For each avocado variety and market, a model was fitted on the first 78 weeks of data and then tested on the remaining 39 weeks to assess performance. 
Unlike in the other experiments where the model was tuned using cross-validation, here we wish to respect the chronological ordering of the observations. 
We therefore train on the first 39 observations, then validate on the next 39 in order to select the model, falling within the set-up of Theorem~\ref{thm:mainresult}.
Once our model is selected we then retrain it on all 78 of the training observations before testing on the hitherto-unseen test set of the last 39 observations. 

The ordering of variables we used with our nested regression scheme was motivated by the following considerations.
For a univariate time series with no seasonal effects, we would typically order the variables by ranking them from most to least recent (with most recent being the most `important'). Here there are effectively 106 time series which are observed weekly, so there may be some seasonal effect.

The ordering used here was constructed by first splitting the time series into groups of decreasing `importance' as follows:
\begin{enumerate}
  \item the particular time series that we are modelling;
  \item the complementary variety to the time series we are modelling (e.g.\ were we modelling Albany-conventional, here we would take Albany-organic); 
  \item everything else.
\end{enumerate}
Within each of these groups, the reading from 52 weeks (one year) previous was first in the ordering, with the rest ordered from newest to oldest.

\begin{figure}
\centering
\includegraphics[width=\textwidth]{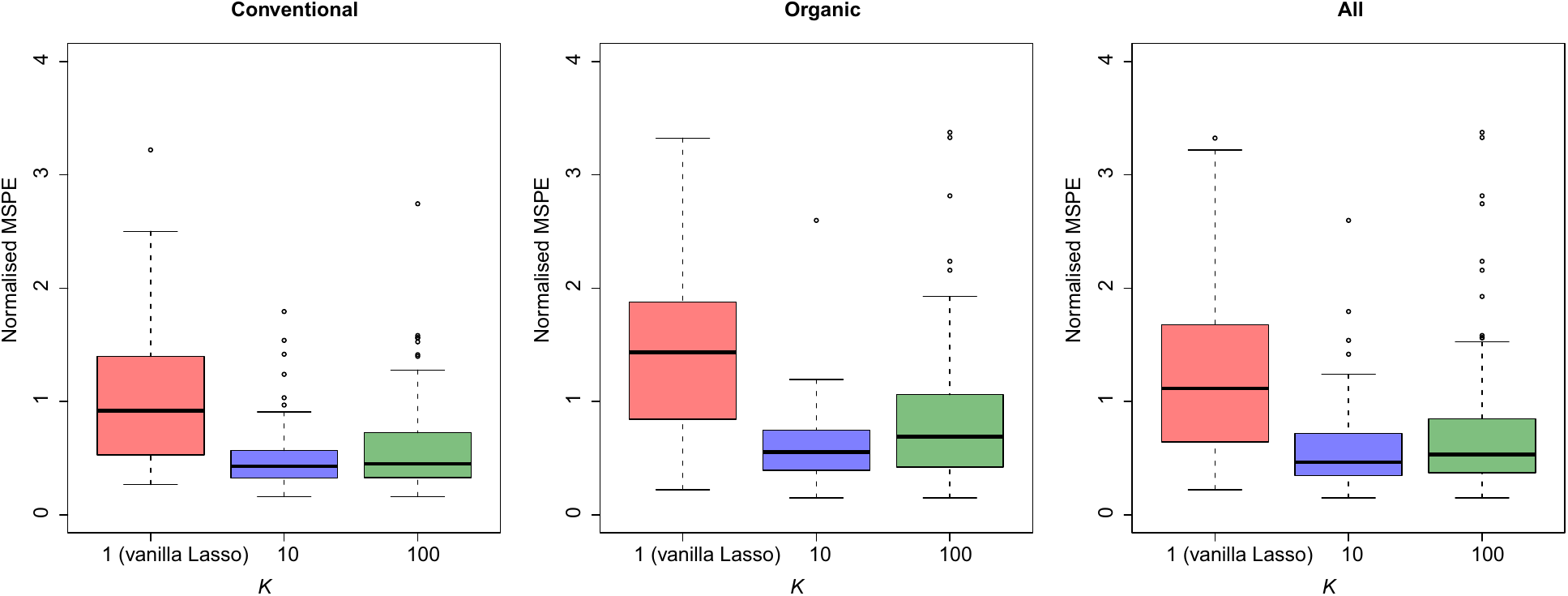}
\caption{Mean squared prediction error (MSPE) across the models for our nested regressions with varying $K$, with the leftmost boxplot in each panel corresponding to the special case of the regular Lasso. Left and centre plots show the differences between the conventional and organic time series, the right plot includes results for both varieties.
\label{fig:avocado}}	
\end{figure}

Each of the response vectors were scaled to have unit variance.
Figure~\ref{fig:avocado} contains mean squared prediction error for each time series, fitted with both vanilla Lasso models, as well as our approach with $K= 10$ and $K=100$. 
In both cases, using our approach substantially improves the quality of the predictions and that $K=10$ gives the largest improvement. 
This illustrates the flexibility of our approach with respect to the origin of the ordering over the variables, and how it can improve prediction performance in a range of scenarios.
The median times for the $106$ regressions were $1.32$s for $K=1$, $1.53$s for $K=10$ and $3.46$s for $K=100$.

\section{Discussion}
In this work we have introduced a simple and general nested regression approach for using potential prior information on the importance of variable in a regression problem. Such prior information may be available in a variety of settings of interest, including time series analysis, settings with missing data, and where some corruption of the data is suspected, among many others.
We have described a simple computational strategy for implementing approach with the Lasso, and ridge regression, and provided in Theorem~\ref{thm:mainresult} some theoretical support for our scheme via a general result on selecting estimators in high-dimensional linear regression settings. It would be of interest to extend this result to encompass generalised linear models, for example, and also aggregation schemes such as stacking \citep{wolpert1992stacked}. Another open question is whether there is a data-dependent choice of the number of subsets $K$ that can be used.

\bibliographystyle{abbrvnat}
\bibliography{Bibliography.bib}

\begin{thebibliography}{35}
\providecommand{\natexlab}[1]{#1}
\providecommand{\url}[1]{\texttt{#1}}
\expandafter\ifx\csname urlstyle\endcsname\relax
  \providecommand{\doi}[1]{doi: #1}\else
  \providecommand{\doi}{doi: \begingroup \urlstyle{rm}\Url}\fi

\bibitem[Abou~Elhamayed(2018)]{abou2018comparative}
S.~H. Abou~Elhamayed.
\newblock Comparative study on different classification techniques for spam
  dataset.
\newblock \emph{International Journal of Computer and Communication
  Engineering}, 7\penalty0 (4), 2018.

\bibitem[Belloni et~al.(2011)Belloni, Chernozhukov, and Wang]{Belloni2011}
A.~Belloni, V.~Chernozhukov, and L.~Wang.
\newblock Square-root lasso: pivotal recovery of sparse signals via conic
  programming.
\newblock \emph{Biometrika}, 98\penalty0 (4):\penalty0 791--806, 2011.

\bibitem[Chetverikov et~al.(2021)Chetverikov, Liao, and
  Chernozhukov]{chetverikov2021cross}
D.~Chetverikov, Z.~Liao, and V.~Chernozhukov.
\newblock On cross-validated lasso in high dimensions.
\newblock \emph{Annal. Stat.(Forthcoming)}, 2021.

\bibitem[Chun and Kele{\c{s}}(2010)]{chun2010sparse}
H.~Chun and S.~Kele{\c{s}}.
\newblock Sparse partial least squares regression for simultaneous dimension
  reduction and variable selection.
\newblock \emph{Journal of the Royal Statistical Society: Series B (Statistical
  Methodology)}, 72\penalty0 (1):\penalty0 3--25, 2010.

\bibitem[Datta et~al.(2017)Datta, Zou, et~al.]{datta2017cocolasso}
A.~Datta, H.~Zou, et~al.
\newblock Cocolasso for high-dimensional error-in-variables regression.
\newblock \emph{Annals of Statistics}, 45\penalty0 (6):\penalty0 2400--2426,
  2017.

\bibitem[Dezeure et~al.(2015)Dezeure, B{\"u}hlmann, Meier, and
  Meinshausen]{dezeure2015high}
R.~Dezeure, P.~B{\"u}hlmann, L.~Meier, and N.~Meinshausen.
\newblock High-dimensional inference: confidence intervals, p-values and
  r-software hdi.
\newblock \emph{Statistical science}, pages 533--558, 2015.

\bibitem[Feng and Yu(2019)]{feng2019restricted}
Y.~Feng and Y.~Yu.
\newblock The restricted consistency property of leave-nv-out cross-validation
  for high-dimensional variable selection.
\newblock \emph{Statistica Sinica}, 29\penalty0 (3):\penalty0 1607--1630, 2019.

\bibitem[Friedman et~al.(2010)Friedman, Hastie, and
  Tibshirani]{friedman2010regularization}
J.~Friedman, T.~Hastie, and R.~Tibshirani.
\newblock Regularization paths for generalized linear models via coordinate
  descent.
\newblock \emph{Journal of Statistical Software}, 33\penalty0 (1):\penalty0
  1--22, 2010.
\newblock URL \url{https://www.jstatsoft.org/v33/i01/}.

\bibitem[Hoerl and Kennard(1970)]{hoerl1970ridge}
A.~E. Hoerl and R.~W. Kennard.
\newblock Ridge regression: Biased estimation for nonorthogonal problems.
\newblock \emph{Technometrics}, 12\penalty0 (1):\penalty0 55--67, 1970.

\bibitem[Kalambe et~al.(2020)Kalambe, Rufus, Karar, and
  Poddar]{kalambe2020descriptor}
S.~S. Kalambe, E.~Rufus, V.~Karar, and S.~Poddar.
\newblock Descriptor-length reduction using low-variance filter for visual
  odometry.
\newblock In \emph{Proceedings of 3rd International Conference on Computer
  Vision and Image Processing}, pages 1--11. Springer, 2020.

\bibitem[Kiggins()]{avocadokaggle}
J.~Kiggins.
\newblock Avocado prices.
\newblock \url{https://www.kaggle.com/neuromusic/avocado-prices}.
\newblock Accessed: 2021-06-04.

\bibitem[Langkun et~al.(2020)Langkun, Sthevanie, and
  Atastina]{langkun2020feature}
E.~R.~M. Langkun, F.~Sthevanie, and I.~Atastina.
\newblock Feature selection on facial expression recognition system using low
  variance filter.
\newblock \emph{eProceedings of Engineering}, 7\penalty0 (2), 2020.

\bibitem[Loh et~al.(2012)Loh, Wainwright, et~al.]{loh2012high}
P.-L. Loh, M.~J. Wainwright, et~al.
\newblock High-dimensional regression with noisy and missing data: Provable
  guarantees with nonconvexity.
\newblock \emph{The Annals of Statistics}, 40\penalty0 (3):\penalty0
  1637--1664, 2012.

\bibitem[Meinshausen(2007)]{meinshausen2007relaxed}
N.~Meinshausen.
\newblock Relaxed lasso.
\newblock \emph{Computational Statistics \& Data Analysis}, 52\penalty0
  (1):\penalty0 374--393, 2007.

\bibitem[Micchelli et~al.(2010)Micchelli, Morales, and
  Pontil]{micchelli2010family}
C.~Micchelli, J.~Morales, and M.~Pontil.
\newblock A family of penalty functions for structured sparsity.
\newblock \emph{Advances in Neural Information Processing Systems},
  23:\penalty0 1612--1623, 2010.

\bibitem[Nardi and Rinaldo(2011)]{nardi2011autoregressive}
Y.~Nardi and A.~Rinaldo.
\newblock Autoregressive process modeling via the lasso procedure.
\newblock \emph{Journal of Multivariate Analysis}, 102\penalty0 (3):\penalty0
  528--549, 2011.

\bibitem[Pitman and Tran(2015)]{pitman2015size}
J.~Pitman and N.~M. Tran.
\newblock Size-biased permutation of a finite sequence with independent and
  identically distributed terms.
\newblock \emph{Bernoulli}, 21\penalty0 (4):\penalty0 2484--2512, 2015.

\bibitem[{R Core Team}(2021)]{R}
{R Core Team}.
\newblock \emph{R: A Language and Environment for Statistical Computing}.
\newblock R Foundation for Statistical Computing, Vienna, Austria, 2021.
\newblock URL \url{https://www.R-project.org/}.

\bibitem[Rosenbaum et~al.(2013)Rosenbaum, Tsybakov,
  et~al.]{rosenbaum2013improved}
M.~Rosenbaum, A.~B. Tsybakov, et~al.
\newblock Improved matrix uncertainty selector.
\newblock In \emph{From Probability to Statistics and Back: High-Dimensional
  Models and Processes--A Festschrift in Honor of Jon A. Wellner}, pages
  276--290. Institute of Mathematical Statistics, 2013.

\bibitem[Saputra et~al.(2018)Saputra, Masputra, Syarif, and
  Ramli]{saputra2018botnet}
F.~A. Saputra, M.~F. Masputra, I.~Syarif, and K.~Ramli.
\newblock Botnet detection in network system through hybrid low variance
  filter, correlation filter and supervised mining process.
\newblock In \emph{2018 Thirteenth International Conference on Digital
  Information Management (ICDIM)}, pages 112--117. IEEE, 2018.

\bibitem[Shah and B{\"u}hlmann(2019)]{shah2019double}
R.~D. Shah and P.~B{\"u}hlmann.
\newblock Double-estimation-friendly inference for high-dimensional
  misspecified models.
\newblock \emph{arXiv preprint arXiv:1909.10828}, 2019.

\bibitem[Shah et~al.(2020)Shah, Frot, Thanei, and Meinshausen]{shah2020right}
R.~D. Shah, B.~Frot, G.-A. Thanei, and N.~Meinshausen.
\newblock Right singular vector projection graphs: fast high dimensional
  covariance matrix estimation under latent confounding.
\newblock \emph{Journal of the Royal Statistical Society: Series B (Statistical
  Methodology)}, 82\penalty0 (2):\penalty0 361--389, 2020.

\bibitem[Silipo et~al.(2014)Silipo, Adae, Hart, and Berthold]{filter2014seven}
R.~Silipo, I.~Adae, A.~Hart, and M.~Berthold.
\newblock Seven techniques for dimensionality reduction.
\newblock Technical report, Technical report, 2014.

\bibitem[Singh et~al.(2017)]{singh2017novel}
S.~Singh et~al.
\newblock A novel algorithm to preprocess cancerous gene expression dataset for
  efficient gene selection.
\newblock In \emph{2017 2nd International Conference for Convergence in
  Technology (I2CT)}, pages 632--635. IEEE, 2017.

\bibitem[Sun and Zhang(2012)]{sun2012scaled}
T.~Sun and C.-H. Zhang.
\newblock Scaled sparse linear regression.
\newblock \emph{Biometrika}, 99\penalty0 (4):\penalty0 879--898, 2012.

\bibitem[Sun and Zhang(2013)]{sun2013sparse}
T.~Sun and C.-H. Zhang.
\newblock Sparse matrix inversion with scaled lasso.
\newblock \emph{The Journal of Machine Learning Research}, 14\penalty0
  (1):\penalty0 3385--3418, 2013.

\bibitem[Tibshirani(1996)]{tibshirani1996regression}
R.~Tibshirani.
\newblock Regression shrinkage and selection via the lasso.
\newblock \emph{Journal of the Royal Statistical Society: Series B
  (Methodological)}, 58\penalty0 (1):\penalty0 267--288, 1996.

\bibitem[Tibshirani and Suo(2016)]{tibshirani2016ordered}
R.~Tibshirani and X.~Suo.
\newblock An ordered lasso and sparse time-lagged regression.
\newblock \emph{Technometrics}, 58\penalty0 (4):\penalty0 415--423, 2016.

\bibitem[Van~de Geer and B{\"u}hlmann(2009)]{van2009conditions}
S.~A. Van~de Geer and P.~B{\"u}hlmann.
\newblock On the conditions used to prove oracle results for the lasso.
\newblock \emph{Electronic Journal of Statistics}, 3:\penalty0 1360--1392,
  2009.

\bibitem[Wolpert(1992)]{wolpert1992stacked}
D.~H. Wolpert.
\newblock Stacked generalization.
\newblock \emph{Neural networks}, 5\penalty0 (2):\penalty0 241--259, 1992.

\bibitem[Yuan and Lin(2006)]{yuan2006model}
M.~Yuan and Y.~Lin.
\newblock Model selection and estimation in regression with grouped variables.
\newblock \emph{Journal of the Royal Statistical Society: Series B (Statistical
  Methodology)}, 68\penalty0 (1):\penalty0 49--67, 2006.

\bibitem[Zhang(2010)]{zhang2010nearly}
C.-H. Zhang.
\newblock Nearly unbiased variable selection under minimax concave penalty.
\newblock \emph{The Annals of statistics}, 38\penalty0 (2):\penalty0 894--942,
  2010.

\bibitem[Zhao and Yu(2006)]{zhao2006model}
P.~Zhao and B.~Yu.
\newblock On model selection consistency of lasso.
\newblock \emph{The Journal of Machine Learning Research}, 7:\penalty0
  2541--2563, 2006.

\bibitem[Zhu et~al.(2019)Zhu, Wang, and Samworth]{zhu2019high}
Z.~Zhu, T.~Wang, and R.~J. Samworth.
\newblock High-dimensional principal component analysis with heterogeneous
  missingness.
\newblock \emph{arXiv preprint arXiv:1906.12125}, 2019.

\bibitem[Zou(2006)]{zou2006adaptive}
H.~Zou.
\newblock The adaptive lasso and its oracle properties.
\newblock \emph{Journal of the American Statistical Association}, 101\penalty0
  (476):\penalty0 1418--1429, 2006.

\end{thebibliography}

\appendix
\section{Proof of Theorem~\ref{thm:mainresult}}
	We begin with the so-called basic inequality
\begin{align}
	\frac{1}{n}\| X ( \hat{\beta} - \beta)\|_2^2 &\leq \frac{1}{n} \| X ( \hat{\beta}^* - \beta ) \|_2^2 + \frac{2}{n} | \varepsilon^T X ( \hat{\beta}^* - \hat{\beta} ) |\label{eq:basicinequality},
\end{align}
which follows from the fact that $\| Y - X \hat{\beta} \|_2^2 \leq \| Y - X \hat{\beta}^{(m)} \|_2^2$ for all $m \in \{ 1, \ldots, M \}$.
We will first control the second term on the right-hand side of \eqref{eq:basicinequality}. 
Since the entries of $\varepsilon$ are independent $\sigma$-sub-Gaussian distributed and independent of $X$, we have that with probability at least $1 - 2M^{-c_2}$,
\begin{align*}
	\frac{2}{n} | \varepsilon^T X(\hat{\beta}^{(m)} - \hat{\beta})| \leq \sigma\sqrt{8(1 + c_2)}\sqrt{\frac{\log M}{n}}\frac{1}{\sqrt{n}}\|X(\hat{\beta}^{(m)} - \hat{\beta})\|_2
\end{align*}
for all $m =1,\ldots,M$.
In particular, since $\hat{\beta}^* \in \{ \hat{\beta}^{(1)}, \ldots, \hat{\beta}^{(M)} \}$, it follows that
\begin{align}
	\frac{1}{n} \|X(\hat{\beta} - \beta )\|_2^2 &\leq \frac{1}{n} \| X (\hat{\beta}^* - \beta ) \|_2^2 +\sigma\sqrt{8(1 + c_2)}\sqrt{\frac{\log M}{n}}\frac{1}{\sqrt{n}}\|X(\hat{\beta} - \hat{\beta}^*)\|_2 \notag \\ 
	&\leq \frac{1}{n} \| X (\hat{\beta}^* - \beta ) \|_2^2 + \sigma\sqrt{8(1 + c_2)}\sqrt{\frac{\log M}{n}} \left( \frac{1}{\sqrt{n}}\|X(\hat{\beta}^* - \beta)\|_2 + \frac{1}{\sqrt{n}} \| X(\hat{\beta} - \beta)\|_2 \right),\label{eq:ineq_line} 
\end{align}
using the triangle inequality in the final line.
Now, we use the observation that for $a,b,c \geq 0$, 
\begin{align*}
	a^2 &\leq b^2 + c(a+b) \\ 
	\implies a^2 - ca &\leq b^2 + cb \\
	\implies 	\left(a - \frac{1}{2}c \right)^2 &\leq 	\left( b + \frac{1}{2}c \right)^2 \\
	\implies a &\leq b + c.
\end{align*}
Application of this to \eqref{eq:ineq_line} yields
\begin{align}
	\frac{1}{\sqrt{n}} \|X(\hat{\beta} - \beta )\|_2 &\leq	\frac{1}{\sqrt{n}} \| X( \hat{\beta}^* - \beta ) \|_2  + \sigma\sqrt{8(1 + c_2)}\sqrt{\frac{\log M}{n}} . \label{eq:second_basic_ineq}
\end{align}
Now let us write $\delta_m := \hat{\beta}^{(m)} -\beta$ and $v_m :=  \Sigma^{1/2} \delta_m / \sqrt{\delta_m^T \Sigma \delta_m}$ so $\|v_m\|_2=1$. Then
\begin{align*}
	\frac{\delta_m^T \left( \Sigma - \frac{1}{n}X^T X\right)\delta_m}{\delta_m^T \Sigma \delta_m} &= v_m^T\left(I_p - \frac{1}{n}W^T w\right)v_m \\ 
	&= 1 - \frac{1}{n}v_m^TW^TWv_m \\  
	&= 1 - \frac{1}{n}\sum_{i=1}^n V_{mi}^2
\end{align*}
where $V_m := W v_m$. Note that each of the $n$ entries of $V_m$ are independent, centred, and sub-Gaussian with variance proxy $\nu^2$ and $\E V_{mi}^2 = 1$.
Recall that for a general random variable $U$ that is centred and $\omega$-sub-Gaussian, $U^2 - \E [ U^2]$ is sub-exponential with parameters $(32 \omega^4, 4 \omega^2)$.
Thus, $ 1 - \frac{1}{n}\sum_{i=1}^n V_{mi}^2$ is a centred sub-exponential random variable with parameters $(\frac{1}{n} 32 \nu^4, \frac{1}{n} 4 \nu^2)$ by the additivity property of independent sub-exponential random variables. Therefore for each $m$,
\begin{align*}
	\pr \left( \left| \frac{\delta_m^T ( \Sigma - \frac{1}{n}X^T X)\delta_m}{\delta_m^T \Sigma \delta_m} \right| \geq t \right) \leq 2 \exp \left( - \frac{tn ( t \wedge 8 \nu^2 )}{64 \nu^4}\right) \end{align*}
and so
\[
\pr  \left( \max_{m \in \{ 1, \ldots, M \} } \left\{ \left| \frac{\delta_m^T ( \Sigma - \frac{1}{n}X^T X)\delta_m}{\delta_m^T \Sigma \delta_m} \right| \right\} \geq t \right) \leq 2 M\exp \left( - \frac{tn ( t \wedge 8 \nu^2 )}{64 \nu^4}\right).
\]
We then have that with probability at least $1 - 2 M^{-c_1}$ for some constant $c_1>0$, 
\begin{align*}
	\left|(\hat{\beta}^{(m)} - \beta)^T\left(\Sigma - \frac{1}{n}X^T X\right)(\hat{\beta}^{(m)} - \beta)\right| \leq 8\nu^2\sqrt{1 + c_1}\sqrt{\frac{\log M}{n}}(\hat{\beta}^{(m)} - \beta)^T\Sigma(\hat{\beta}^{(m)} - \beta)
\end{align*}
for all $m=1,\ldots,M$.
In the above we have used the assumption that $c_1 + 1 < n /  \log M$ which implies that $(t \wedge 8 \nu^2) = t$ for our choice of $t$.
On this event it follows that for each $m$, 
\begin{align*}
	\left| \frac{1}{\sqrt{n}} \|X ( \hat{\beta}^{(m)} -  \beta)\|_2 - \| \Sigma^{1/2} ( \hat{\beta}^{(m)} - \beta ) \|_2 \right| &\leq  2 \sqrt{2}\nu(1 + c_1)^{1/4}\left(\frac{\log M}{n}\right)^{1/4} \| \Sigma^{1/2}( \hat{\beta}^{(m)} - \beta) \|_2,
\end{align*}
which gives the following inequalities
\begin{align*}
	\frac{1}{\sqrt{n}} \| X( \hat{\beta} - \beta) \|_2 &\geq \left( 1 - 2 \sqrt{2} \nu (1 + c_1)^{1/4} \left( \frac{\log M}{n} \right)^{1/4} \right) \| \Sigma^{1/2} ( \hat{\beta} - \beta) \|_2, \\
	\frac{1}{\sqrt{n}} \| X( \hat{\beta}^* - \beta) \|_2 &\leq \left( 1 + 2 \sqrt{2} \nu (1 + c_1)^{1/4} \left( \frac{\log M}{n} \right)^{1/4} \right) \| \Sigma^{1/2} ( \hat{\beta}^* - \beta) \|_2.
\end{align*}
Combining these with \eqref{eq:second_basic_ineq} gives that with probability at least $1 - 2M^{-c_1} - 2M^{-c_2}$, 
\begin{align*}
	\| \Sigma^{1/2} ( \hat{\beta} - \beta ) \|_2 \leq \frac{1 + \Psi}{1 - \Psi}  \| \Sigma^{1/2}(\hat{\beta}^* - \beta ) \|_2  + \frac{1}{1 - \Psi}2\sqrt{2}\sigma \sqrt{1 + c_2} \sqrt{\frac{\log M}{n}}
\end{align*} 
where $\Psi = 2\sqrt{2}\nu(1 + c_1)^{1/4} ( (\log M) / n )^{1/4}$, as required. 
\qed
\end{document}